\documentclass[aps,prl, twocolumn,superscriptaddress]{revtex4-1}
\usepackage{graphicx,times}
\usepackage{dcolumn}
\usepackage{bm}
\usepackage{lmodern}
\usepackage{mathtools}
\usepackage{dsfont}
\usepackage{mathrsfs}
\usepackage{amsthm,amsfonts,mathtools,amsmath,amssymb}
\usepackage{mathbbol}
\usepackage{hyperref}
\usepackage{xcolor}
\usepackage{url}
\usepackage{microtype}

\newcommand{\rrho}{\hat{\boldsymbol{\rho}}}

\newcommand{\Tr}{\mathrm{Tr}}
\newcommand{\Id}{\mathds{I}}

\newcommand{\Var}{\mathrm{Var}}
\theoremstyle{plain}

\theoremstyle{definition}

\newcommand{\mga}[1]{{#1}}

\begin{document}


\title{Joint estimation of position and momentum with arbitrarily high precision using non-Gaussian states}

\author{Massimo Frigerio}
\email{Electronic address: massimo.frigerio@lkb.upmc.fr}
\affiliation{Laboratoire Kastler Brossel, Sorbonne Universit\'{e}, CNRS, ENS-Universit\'{e} PSL,  Coll\`{e}ge de France, 4 place Jussieu, F-75252 Paris, France}

\author{Matteo G. A. Paris}
\email{Electronic address: matteo.paris@fisica.unimi.it}
\affiliation{Dipartimento di Fisica  ``Aldo Pontremoli'',
Universit\`a degli 
Studi di Milano, I-20133 Milano, Italy}

\author{Carlos E. Lopetegui}
\email{Electronic address: \allowbreak{} carlos-ernesto.lopetegui-gonzalez@lkb.upmc.fr}
\affiliation{Laboratoire Kastler Brossel, Sorbonne Universit\'{e}, CNRS, ENS-Universit\'{e} PSL,  Coll\`{e}ge de France, 4 place Jussieu, F-75252 Paris, France}

\author{Mattia Walschaers}
\email{Electronic address: mattia.walschaers@lkb.upmc.fr}
\affiliation{Laboratoire Kastler Brossel, Sorbonne Universit\'{e}, CNRS, ENS-Universit\'{e} PSL,  Coll\`{e}ge de France, 4 place Jussieu, F-75252 Paris, France}

\date{\today} 

\begin{abstract}
\end{abstract}

\begin{abstract}
We address the joint estimation of changes in the position and linear momentum of a quantum particle—or, equivalently, changes in the complex field of a bosonic mode. Although these changes are generated by non-commuting operators, we show that leveraging non-Gaussianity enables their simultaneous estimation with arbitrarily high precision and arbitrarily low quantum incompatibility. Specifically, we demonstrate that any pure non-Gaussian state provides an advantage over all Gaussian states, whether pure or mixed. Moreover, properly tuned non-Gaussian mixtures of Gaussian states can also serve as a resource.
\end{abstract}
\maketitle
{\em Introduction---}
Uncertainty relations  establish a lower bound to the product of the variances of the conjugated position and momentum observables on any quantum state, thereby implying that one cannot assign both position and momentum values with arbitrarily high precision to any quantum system. Nonetheless, it has long been acknowledged that it is possible to simultaneously estimate a position and a momentum shift \cite{Duivenvoorden2017}, e.g. by using compass states \cite{Zurek2001,Oh2020} or grid states \cite{PhysRevA.64.012310}. Such shifts are generated by the corresponding conjugate momentum and position operators, respectively. This estimation problem is fully equivalent to simultaneous intensity and phase estimation in quantum optics, but it can also be applied in other platforms such as superconducting microwave cavities \cite{walker1986multiport,Deng2024} and trapped ions \cite{fluhmann_encoding_2019}. 
 
 From the phase-space viewpoint, it is intuitively clear that Gaussian probe states have limited powers in this task, since they cannot be arbitrarily narrow with respect to two conjugate observables (or field-quadratures, in quantum optics). As a matter of fact, all the states that have been put forward for simultaneous estimation of displacements in conjugate quadratures are non-Gaussian. However, no general result concerning the relation between non-Gaussianity and this intrinsically quantum, multi-parameter estimation problem has been put forward. In this work, we formalize this intuition in the form of a strict lower bound on the quantum incompatibility between position and momentum shifts on Gaussian states, using the geometrical approach to quantum estimation theory and the notion of Uhlmann curvature \cite{Uhlmann1986,Campbell1986,Bengtsson2006,Suzuki2019,Carollo2019}. Leveraging on these tools from multi-parameter quantum metrology, we prove that \emph{any} pure non-Gaussian probe state provides an advantage in this joint estimation task. In particular, we show that arbitrarily high precision can be achieved for both parameters simultaneously, albeit not with a Heisenberg-type scaling with respect to the probe's energy.  In doing this, we explicitly show that Fock states are optimal. Since it is, in principle, possible that the probe state is sensitive only to a single joint function of the two parameters and not to each of them independently, we also take the \emph{sloppiness} \cite{brown2003statistical,brown2004statistical,PhysRevLett.97.150601,machta2013parameter,Goldberg21,Yang23,Frigerio2024,PhysRevA.111.012414,hassani2024privacy} of the model into consideration. 
 
 Our result provides a rigorous example of a quantum estimation task where some form of non-Gaussianity is resourceful with respect to all Gaussian states. Crucially, it also relies on the observation that any pure quantum state with high variances with respect to two conjugate quadratures must be highly structured in phase-space and highly non-Gaussian. In this sense, our result can also be seen as another no-go results for Gaussian states in quantum information theory \cite{PhysRevLett.89.137903,PhysRevLett.89.137904,PhysRevResearch.5.L032010, PhysRevLett.88.097904}. Finally, in contrast to computational tasks \cite{PhysRevLett.109.230503,Veitch_2012,PhysRevLett.130.090602}, we find that Wigner negativity and quantum non-Gaussianity are neither necessary nor sufficient for reaching an advantage in this estimation task. \\

{\em Estimating position and momentum shifts with Gaussian states---}
Consider a generic pure state $\vert \psi \rangle$ for a quantum particle in one-dimensional real space, so that the reference Hilbert space is $\mathcal{H} = \mathrm{L}^2 ( \mathds{R} )$, the space of complex-valued square-integrable functions on the real line modulo equivalence almost everywhere. The particle is subject to a sudden change in position by an amount $x_0$ in the positive direction\footnote{But $x_0$ can also be negative}, encoded by the operator $e^{- i x_{0} \hat{p}}$, where $\hat{p} = - \frac{i}{\hbar} \partial_x$ is the linear momentum operator, followed by a velocity boost, encoded by $e^{- i p_{0} \hat{x}}$, which simply acts as a local and linear phase change in the position eigenbasis. We assume that non-adiabatic effects resulting from the sudden application of these operations can be neglected, effectively considering the unitary operators that we introduced as a correct description of the particle state $\vert \phi (x_0, p_0) \rangle$ after the processes:
\begin{equation}
    \vert \phi (x_0, p_0) \rangle \ = \ e^{- i p_{0} \hat{x}} e^{- i x_{0} \hat{p}} \vert \psi \rangle
\end{equation}
 We now seek to estimate at the same time the values of $x_0$ and $p_0$ with the highest precision allowed by quantum mechanics and then see how their uncertainties change as a function of the initial probe state $\vert \psi \rangle$. To this end, we can start by computing the Symmetric Logarithmic Derivative - Quantum Fisher Information (SLD-QFI) matrix\footnote{We refer to the Supplemental Material for the definitions.} for these two parameters. Notice that if we applied the two operations in the opposite order, the result would change just by an overall phase shift, which depends on both $x_0$ and $p_0$ but is constant in position, therefore it is immaterial. As a consequence, also the QFI matrix will be the same. 
 
 For Gaussian models, the QFI matrix is given by \cite{Safranek2018,Safranek17}
\begin{equation}
\label{eq:QFIGauss}
    \boldsymbol{\mathcal{Q}} [ \rrho_{G} ] \ = \ \sigma^{-1}_{\rrho_{G}}
\end{equation}
which is also valid for mixed Gaussian states. Here $\sigma_{\rrho}$ is the \emph{covariance matrix} (CM) of the position and momentum variables in the state $\rrho$, defined as $[\sigma_{\rrho} ]_{jk} = \frac{1}{2} \Tr[\rrho (\hat{R}_j \hat{R}_k + \hat{R}_k \hat{R}_j ) - \Tr[\rrho \hat{R}_j] \Tr[\rrho \hat{R}_k ]$ 
 with $\hat{R}_1 = \hat{x}$, $\hat{R}_2 = \hat{p}$ and $j,k \in \{1,2\}$ and $[\hat{R}_{j} , \hat{R}_{k} ] = i \Omega_{jk}$ where:
 \begin{equation}
     \Omega = \left( \begin{array}{cc} 0 & 1 \\ -1  & 0  \end{array}   \right) 
 \end{equation}
 is the symplectic form \footnote{With our conventions, the covariance matrix of the vacuum state is $\sigma_0 = \frac{1}{2} \Id$}. For pure Gaussian states $\lvert \psi_G\rangle$, in particular, $\sigma_{\psi_G}^{-1} = 2 \Omega^{T} \sigma_{\psi_G} \Omega$, as one can prove from the equality $\sigma_{\psi_G} = \frac{1}{2} S S^{T}$, where $S$ is a symplectic matrix. The key observation is that for Gaussian pure states $\det \sigma_{\psi_G} = \frac{1}{4}$ and the two eigenvalues of $\boldsymbol{\mathcal{Q}}$ cannot both be arbitrarily high: this provides a first clear example where the power of Gaussian states is limited for quantum metrology applications. In general, moreover, for mixed Gaussian states of a single mode one has $\sigma_{\rrho_G} = \frac{1}{2 \mu} S S^{T}$, where $\mu = \Tr[ \rrho_{G}^{2} ]$ is the purity, therefore $\det \mathcal{Q} [ \rrho_{G} ] = 4 \mu^{2}$ decreases even further for mixed Gaussian states and the model becomes sloppier. 
 
 On the other hand, the SLD-QFI Cramér-Rao bound in the multiparameter case is not always tight, since in general there is no single POVM that can be used to estimate both parameters simultaneously at the level of precision entailed by the QFI matrix. To address this issue, we must compute the R-quantity
 \cite{Carollo2019,Razavian20}, which bounds the difference between the SLD-QFI Cramér-Rao bound and the Holevo bound (see the Supplemental material), which is instead tight. For two parameters, we have that:
 \begin{equation}
     \mathcal{R} \ = \ \sqrt{ \dfrac{ \det  2 \bm{\mathcal{U}}}{\det \bm{\mathcal{Q}}} }
 \end{equation}
 where the anti-symmetric matrix $\bm{\mathcal{U}}$, usually referred to as the Uhlmann curvature, is defined as:
\begin{equation}
   [ \bm{\mathcal{U}}]_{\alpha \beta} \ := \  \frac{i}{4} \Tr \left[ \rrho \big[ \hat{\mathrm{L}}_{\alpha} , \hat{\mathrm{L}}_{\beta} \big] \right]
\end{equation}
and $\hat{\mathrm{L}}_{\alpha}$ is the symmetric logarithmic derivative operator for the parameter indexed by $\alpha$\footnote{We fixed $x_0$ for $\alpha=1$ and $x_0$ for $\alpha= 2$}. Thus, analogously to a Berry phase, it is the imaginary part of the \emph{quantum geometric tensor} of the quantum statistical model at hand. In general $0 \leq \mathcal{R} \leq 1$ and, when it is zero, the SLD-QFI bound coincides with the Holevo bound and therefore is guaranteed to be asymptotically achievable \footnote{One might need to perform collective measurements on asymptotically large number of copies of the state encoding the parameters, though}. In the case of a pure state model $\{ \vert \phi \rangle \}$, the Uhlmann curvature can be rewritten as:
\begin{equation}
    [\bm{\mathcal{U}}]_{\alpha \beta} \ = \ - 2   \mathrm{Im} \left\{ \langle \partial_\alpha \phi \vert \partial_\beta \phi \rangle - \langle \phi \vert \partial_\alpha \phi \rangle \langle \partial_\beta \phi \vert \phi \rangle \right\}
\end{equation} 
Since the quantity inside the brackets becomes real when $\alpha = \beta$, the diagonal elements of the Uhlmann curvature are always zero. Moreover, it is apparent that $[\bm{\mathcal{U}}]_{\alpha \beta} = - [\bm{\mathcal{U}}]_{\beta \alpha}$, therefore we only have one entry to compute in the case of two parameters.  In the case we are studying, we find that $ 2 \mathrm{Im}  \langle \partial_{p_0} \phi \vert \partial_{x_0} \phi \rangle =  \mathrm{Im} \langle \psi \vert [ \hat{x} + x_0, \hat{p} ] \vert \psi \rangle = 1$, which is constant and independent of the probe state $\vert \psi \rangle$, by virtue of the uncertainty relation between $\hat{x}$ and $\hat{p}$. Ultimately, we find $ \bm{\mathcal{U}} \ = \  \Omega$,
therefore $\det  2 \bm{\mathcal{U}} = 4$ and since $\det  \bm{\mathcal{Q}} = 4$ for a pure Gaussian model, we always have $\mathcal{R}=1$. 
We must however consider separately the issue of quantum incompatibility for mixed Gaussian states: for Gaussian models and displacement parameters, it is possible to derive the identity $\boldsymbol{\mathcal{U}} = \frac{1}{4} \sigma^{-1} \Omega \sigma^{-1} =   \mu^2  \Omega$ (see Eq.(18) of \cite{Safranek2018}) and the quantum incompatibility parameter becomes equal to the purity: $\mathcal{R}_{G} = \sqrt{\frac{\det [2  \mu^2  \Omega]}{ 4 \mu^2 }} =  \mu$. In practice, this means that Gaussian states with very low purity do not suffer from actual quantum incompatibility (low values of $\mathcal{R}$), but rather they just have low overall sensitivity to displacements (as testified by the small values of the determinant of the QFI).\\

{\em Non-Gaussianity as a resource---}
The observations made so far suggest to look first at the case of non-Gaussian pure states. Applying the standard formula for the SLD-QFI matrix of a pure model, we find:
 \begin{equation}
 \label{eq:QFIpure}
 \begin{aligned}
     \bm{\mathcal{Q}} [ \vert \psi \rangle ]_{\alpha \beta} \ &= \ 4 \mathrm{Re} \left[ \langle \partial_\beta  \phi  \vert \partial_\alpha \phi \rangle  -   \langle   \phi  \vert \partial_\alpha \phi \rangle   \langle \partial_\beta  \phi  \vert \phi \rangle   \right] \ = \\
     & = \ 4 [  \Omega^{T} \sigma_{\psi} \Omega]_{\alpha \beta}
\end{aligned}
 \end{equation}
 The diagonal elements of $\bm{\mathcal{Q}}$  provide lower bounds to the variances of any estimators $\tilde{x}_0$ and $\tilde{p}_0$ of $x_0$ and $p_0$, respectively; they are equal to the variances in momentum and position of the particle in the probe state $\vert \psi \rangle$, $\Delta_{\psi} \hat{p}^2 = \mathcal{Q}_{11}$ and $\Delta_{\psi} \hat{x}^2 = \mathcal{Q}_{22}$ respectively. This result might sound counterintuitive: although the uncertainty relations prevents $\Delta_{\psi} \hat{x}^2$ and $\Delta_{\psi} \hat{p}^2$ from simultaneously attaining arbitrarily low values, it does not prevent these quantities from being arbitrarily large. Yet, arbitrarily large values of these quantities  are associated with arbitrary precision in the estimation of $x_0$ and $p_0$. By noticing that for pure non-Gaussian states $\det \sigma_{\psi} > 1/4$, we conclude that among pure states of a single mode, all non-Gaussian states are resourceful over all Gaussian ones in the problem of jointly estimating two displacements along canonically conjugate quadratures. Even more, $\det \sigma_{\psi}$ is related to an entropic measure of non-Gaussianity \cite{Genoni10}, therefore the behaviour is monotonic and quantitative. Notice also that in Eq.(\ref{eq:QFIGauss}), which is valid for \emph{all} Gaussian states, the CM is inverted, which might seem confusing when compared with the result of Eq.(\ref{eq:QFIpure}); the apparent contradiction is resolved by recalling the inversion formula for the CM of pure Gaussian states mentioned before. 
 
 Let us now look at the quantum incompatibility parameter for pure states; we have $\det \bm{\mathcal{Q}}[\lvert \psi\rangle] = 16 \det \sigma_\psi$ and $\mathcal{U} = \Omega$ as found before, thus we arrive at the general result:
\begin{equation}
    \mathcal{R}_{\text{pure}} \ = \ \sqrt{ \dfrac{1}{4 \det \sigma_{\psi} } } .
\end{equation}
This means that when $\det \sigma_{\psi}$ gets large enough, the $\mathcal{R}$-parameter becomes small and the SLD-QFI bound can almost be saturated. Moreover, $\det \bm{\mathcal{Q}} = 16 \det \sigma_{\psi}$ determines the \emph{sloppiness} of the quantum statistical model: the higher it is, the more statistically independent the two parameters are, meaning that one can estimate their values individually and not just a single function of them. However, by the same token as before, we again stress that this requires a non-Gaussian probe state and any Gaussian pure state will lead to $\mathcal{R} = 1$. Overall, we conclude that the quantum incompatibility parameter $\mathcal{R}$ monotonically decreases with non-Gaussianity and the determinant of the QFI increases proportionally.


Now that we established that only non-Gaussian states allow us to jointly estimate (small value of $\mathcal{R}$) position and momentum shifts at high precision (high values of $\det \bm{\mathcal{Q}}$), it is natural to wonder whether more exotic properties such as quantum non-Gaussianity or Wigner negativity are necessary to give rise to such capabilities. Therefore, the next case to be examined concerns non-Gaussian mixtures of Gaussian states, to see whether they can attain $\mathcal{R} < 1$ and, at the same time, $\det \boldsymbol{\mathcal{Q}} > 4$. The answer is affirmative and an example is provided by the balanced mixture of two squeezed vacuum states \cite{PRApPark22}, one along the $\hat{x}$ quadrature and the other along the $\hat{p}$ quadrature, which by definition does not exhibit any quantum non-Gaussianity:
\begin{equation}
    \rrho_{\hbox{\tiny CNG}}(r) \ = \ \frac{1}{2} \Big( \vert r \rangle \langle r \vert + \vert-r \rangle \langle -r \vert \Big)
\end{equation}
where $\vert r \rangle = \frac {1}{\sqrt {\cosh r}}\sum _{n=0}^{\infty } {\frac {\sqrt {(2n)!}}{2^{n}n!}} \tanh^n r |2n\rangle $ is a single-mode squeezed vacuum state. In the limit of very large $r$, the two states in the mixture are orthogonal and one is very sensitive to displacement in $\hat{p}$ while the second is very sensitive to displacements in $\hat{x}$. One can show that, in this limit, $\boldsymbol{\mathcal{Q}}_{\rrho_{NG}} \approx \frac{1}{2} (\boldsymbol{\mathcal{Q}}_{\vert r \rangle } + \boldsymbol{\mathcal{Q}}_{\vert -r \rangle })$ and $\det \boldsymbol{\mathcal{Q}}_{\rrho_{NG}} = 4 \cosh^{2} 2r$, while $\boldsymbol{\mathcal{U}} = \Omega$ as for the pure states. In particular, not only can arbitrary precision in both parameters be achieved in this case, but the quantum incompatibility decreases exponentially with $r$, since $\mathcal{R} = (\cosh  2r)^{-1}$. In the appendix, we discuss more general cases showing that $\det 2 \boldsymbol{\mathcal{U}}$ can also decrease for unbalanced mixtures. \\

{\em Optimal states at fixed energy---}
Let us now consider the case of a Fock state $\vert n \rangle$. It is well-known that $\sigma_{n} = (n + \frac{1}{2}) \Id_{2}$, so that the variances of both $\hat{x}$ and $\hat{p}$ grow linearly with $n$, while $\det \sigma_n$ grows like $n^2$. Consequently $\mathcal{R}$ decreases as $n^{-1}$, meaning that a high energy Fock state permits to estimate with a very high precision both $x_0$ and $p_0$ without issues of quantum nor classical incompatibility. In particular, the corresponding quantum Cramér-Rao bound, which can be essentially saturated when $\mathcal{R}$ is sufficiently small, will tell us that $\Var (\tilde{x}_0) = \Var (\tilde{p}_0)  \geq \frac{1}{4n + 2}$, leading to the surprising result:
\begin{equation}
    \Var (\tilde{x}_0)  \Var (\tilde{p}_0) \   = \ \frac{1}{(4n + 2)^2} \  \ll \  \frac{1}{4}  \ \ \ (n \gg 0)
\end{equation}
which is suggestive of a violation of the uncertainty relations. Of course, the actual uncertainty relations are never violated, but this simple example shows the subtleties in the difference between computing the variance of non-commuting \emph{observables} and estimating \emph{parameters} encoded by non-commuting generators. 

One might then wonder if Fock states are in a sense optimal, or if one can have a better scaling of the variances of $\hat{x}$ and $\hat{p}$ (simultaneously) with the input energy. In \emph{Supplemental Material} we show that it is impossible to achieve Heisenberg scaling in the estimation of $x_0$ and $p_0$ (even separately) because of energy constraints, thereby also showing that Fock states are the best probe states at fixed energy for this two-parameter quantum estimation problem.

As non-Gaussian mixtures of Gaussian states allow us to outperform all Gaussian states, we should investigate whether mixing Fock states can provide any further metrological advantage. The following identity can be straightforwardly derived from standard results on the SLD-QFI matrix expressed in terms of the diagonal form of the probe state (with $p_{N+1} = 0$):
\begin{widetext}
\begin{equation}
\label{eq:QFIFockmixture}
    \boldsymbol{\mathcal{Q}}\left[ \sum_{n=0}^{N} p_{n} \vert n \rangle \langle n \vert \right] \ = \ 2  \sum_{n=0}^{N} \left\{ \big( 2n + 1 \big) p_n   \ - \ 4( n +1) \frac{p_{n} p_{n+1}}{p_{n} + p_{n+1}}  \right\} \Id_{2}
\end{equation}
\end{widetext}
and where we assumed to have at most $N$ quanta in the mixture\footnote{Note that the formula is valid also for $N \to +\infty$, i.e. for full-rank states}. Clearly, this expression is maximized if only Fock states with the same parity appear in the sum, so that $p_{n} p_{n+1} = 0$. In this case we simply find that the SLD-QFI is proportional to the average number of quanta $\boldsymbol{\mathcal{Q}} =  (4n+2) \Id_{2} \leq (4N +2) \Id_{2}$ so that the mixture is always less sensitive to displacements than the highest Fock state that it contains.\\

{\em Beyond non-Gaussianity---}
As a final question, one may wonder whether quantum non-Gaussianity \cite{MWnG2021} or Wigner negativity are {\em sufficient} to lead to an increased sensitivity with respect to all Gaussian states. To provide and answer, consider the mixed state $\rrho_{\lambda} = (1-\lambda) \vert 0 \rangle \langle 0 \vert + \lambda \vert 1 \rangle \langle 1 \vert $. This state is Wigner negative for $\lambda> \frac{1}{2}$ and quantum non-Gaussian for $\lambda \gtrapprox  0.476$ \cite{MWnG2021}. Using Eq.(\ref{eq:QFIFockmixture}) for mixtures of Fock states, we see that:
\begin{equation}
   \boldsymbol{\mathcal{Q}} [ \rrho_{\lambda}] \ = \  2(1 - 2 \lambda + 4  \lambda^2 ) \Id_{2}
\end{equation}
It saturates the Gaussian limit, $\mathcal{Q} = 2 \Id_{2}$, both for $\lambda = 0$ (vacuum state) and for $\lambda = \frac{1}{2}$ (balanced mixture), while its minimum is for $\lambda = \frac{1}{4}$ for which $\boldsymbol{\mathcal{Q}}_{\lambda} = \frac{3}{2} \Id_{2}$. This rules out the sufficiency of quantum non-Gaussianity, but not that of Wigner negativity, since we get $\det \boldsymbol{\mathcal{Q}} > 4$ precisely when the probe state is Wigner negative. 

Finally, let us consider a photon-added thermal state, which is diagonal in the Fock basis with (normalized) probabilities given by $p^{+,\text{th}}_{n}(\lambda) = n (1-\lambda)^{2} \lambda^{n-1}$, where $\lambda = \frac{\overline{n}_{th}}{\overline{n}_{th}+1}$ and $\overline{n}_{th}$ is the average number of thermal photons in the original thermal state. The Wigner functions of these states always attain negative values at $(x,p)=(0,0)$ in phase-space \cite{PhysRevLett.119.183601}. However, applying Eq.(\ref{eq:QFIFockmixture}), one can check that $\boldsymbol{\mathcal{Q}} \simeq 1.070 \ \Id_{2}$ for $\lambda = 1/2$, a much lower value compared to the vacuum state. We can therefore conclude that Wigner negativity is also not sufficient to ensure a metrological advantage. \\

{\em Conclusions and outlooks---}
We have shown that simultaneous quantum estimation of position and momentum shifts on a single mode can be performed with arbitrarily high precision and  arbitrarily low quantum incompatibility. To this aim, non-Gaussian probes are needed, \emph{all} of which provide an advantage in this specific task when they are pure, as they always increase the classical independence of the two parameters, as measured by the determinant of the QFI matrix. Our result encompasses a number of previous theoretical and experimental works focusing on specific non-Gaussian probe states and specific platforms for this estimation problem \cite{Zurek2001,Duivenvoorden2017,Oh2020,PRApPark22,Deng2024,Valahu24}, while also strengthening the case for the need of non-Gaussianity in estimation tasks. It is also remarkable that non-Gaussianity emerges as a requirement from a quantum statistical model where the parameters are encoded by the simplest Gaussian unitary operations, i.e. displacements. Notice that we have shown that necessity of non-Gaussianity holds for estimation tasks involving a \emph{single mode}, whereas \emph{entangled} Gaussian states of two modes can be employed to obtain the same results, at the cost of having to perform joint measurements of two modes. The trade-off between this entanglement-relying strategy and non-Gaussianity is probably best judged depending on the exact platform and application, but the overall idea of going to two modes is easily understood: 
using pure states of a two-mode system, there is {\em more room in the Hilbert space} \cite{Candeloro24}, and even if the two displacements are encoded on only one of them (say mode $1$), the SLD-QFI matrix is still equal to the matrix of covariances of the generators with respect to the overall pure state. 
This is in turn equal to the CM of mode $1$, which now is not bounded to have determinant equal to $1$, since the state of the first mode does not need to be pure. The argument for non-Gaussianity as a resource is very similar: for every CM we can find a single-mode non-Gaussian ``purification'' (see Supplementary Material). This gives us single-mode pure states with determinant greater than $1$, which leads to a quantum advantage in the estimation task. Here we effectively increase the accessible Hilbert space by dropping the constraint of Gaussianity.

{\em Acknowledgements---}
MF, CEL, and MW acknowledge insightful discussions with Nicolas Treps. 
\mga{MGAP thanks Stefano Olivares, Marco G. Genoni, and Alessandro Ferraro for useful discussions.} MF and MW received financial support through the ANR JCJC project NoRdiC (ANR-21-CE47-0005) and the QuantERA II Project SPARQL that has received funding from the European Union's Horizon 2020 research and innovation programme under Grant Agreement No. 101017733.

\bibliography{main_nGforMPQM}

\end{document}


\title{Supplementary Material for\\ ``non-Gaussianity as a resource in multiparameter quantum metrology''}

\author{Massimo Frigerio}
\email{Electronic address: massimo.frigerio@lkb.upmc.fr}
\affiliation{Laboratoire Kastler Brossel, Sorbonne Universit\'{e}, CNRS, ENS-Universit\'{e} PSL,  Coll\`{e}ge de France, 4 place Jussieu, F-75252 Paris, France}

\author{Mattia Walschaers}
\email{Electronic address: mattia.walschaers@lkb.upmc.fr}
\affiliation{Laboratoire Kastler Brossel, Sorbonne Universit\'{e}, CNRS, ENS-Universit\'{e} PSL,  Coll\`{e}ge de France, 4 place Jussieu, F-75252 Paris, France}

\author{Carlos E. Lopetegui}
\email{Electronic address: carlos-ernesto.lopetegui-gonzalez @lkb.upmc.fr}
\affiliation{Laboratoire Kastler Brossel, Sorbonne Universit\'{e}, CNRS, ENS-Universit\'{e} PSL,  Coll\`{e}ge de France, 4 place Jussieu, F-75252 Paris, France}

\author{Matteo G. A. Paris}
\email{Electronic address: matteo.paris@fisica.unimi.it}
\affiliation{Quantum Technology Lab $\&$ Applied Quantum Mechanics Group, Dipartimento di Fisica  ``Aldo Pontremoli'',
Universit\`a degli 
Studi di Milano, I-20133 Milano, Italy}
\affiliation{INFN, Sezione di Milano, I-20133 Milano, Italy}

\date{\today}
\maketitle
\section{Vademecum of multi-parameter quantum metrology}

In the classical estimation theory of continuous parameters, a classical statistical model is a family of probability density functions $p_{\vec{\lambda}}: X \rightarrow [0,1]$ with respect to some (continuous or discrete) variable $x \in X$, labelled by real parameters $\vec{\lambda}$ belonging to an open subset $\Lambda \subseteq \reals^n$ \cite{LSeveso2020}. Estimation theory aims at reconstructing the true values of the parameters, thus the true probability density function inside the family, starting from a finite sample of $M$ measured outcomes for the random variable and using some estimator $\Theta^{(M)}: X^{M} \to \Lambda$, i.e. a function of the dataset that gives the string of estimated parameters as the output. If this estimator is unbiased, meaning that its expectation value coincides with the true value of the parameters vector:
\[  \mathrm{E}_{\vec{\lambda}} \left[  \Theta^{(M)} \right] = \vec{\lambda}  \]
then the covariance matrix of the parameters estimated with it:
\[ \Sigma_{jk} ( \Theta^{(M)} ) \ = \ \mathrm{E}_{\vec{\lambda}} \left[ \left( ( \Theta^{(M)})_{j} - \lambda_{j} \right)   \left( ( \Theta^{(M)})_{k} - \lambda_{k} \right)        \right] \]
is bounded by the \emph{Cramér-Rao bound}:
\begin{equation}
    \label{eq:CramerRao}
    \Sigma \left( \Theta^{(M)} \right) \ \ \geq \ \ \ \frac{1}{M} \ \mathscr{F}^{-1} [  \vec{\lambda}^{*}]
\end{equation}
where the functional $\mathscr{F}$, known as the \emph{Fisher Information Matrix}, associates a positive-definite invertible matrix to each probability distribution in the statistical model, computed as:

\begin{equation}
    \mathscr{F}_{jk} [ \vec{\lambda} ] \ = \  \int_{X} \ p_{\vec{\lambda}} \  \partial_{{j}}  \left[ \log p_{\vec{\lambda}} \right]      \ \partial_{{k}} \left[ \log p_{\vec{\lambda}} \right] \  dx
\end{equation}
using the short-hand notations $p_{\vec{\lambda}} \equiv p_{\vec{\lambda}} (x)$ and $ \partial_{{j}} \equiv  \partial_{\lambda_{j}} $. 
An estimator is called \emph{asymptotically efficient} if it saturates the Cramér-Rao inequality as $M \to +\infty$ and it always exists for classical statistical models (e.g. the maximum-likelihood estimator and by the Bayesian estimator). 

We proceed by defining a \emph{quantum statistical model} \cite{LSeveso2020} as a map from the open set $\Lambda \subseteq \reals^n$ of real parameters to quantum states on some Hilbert space $\mathcal{H}$, $\vec{\lambda} \mapsto \rrho_{\vec{\lambda}} \in \mathcal{T}( \mathcal{H} ) $, where $\mathcal{T}$ denotes the set of trace-class, bounded, positive-semidefinite linear operators. As long as the map is continuous, $\Lambda$ is open and the operators in the model are all true quantum states, it is possible to define a quantity, \emph{Quantum Fisher Information} (QFI) matrix, that is independent on the specific choice of a measurement:
\begin{equation}
\label{eq:SLDtoQFI}
  \mathcal{Q}_{jk} ( \vec{\lambda} ) \ = \ \Tr \left[ \rrho_{\vec{\lambda}} \,\dfrac{\hat{\mathrm{L}}^{S}_{j} \hat{\mathrm{L}}^{S}_{k} + \hat{\mathrm{L}}^{S}_{k} \hat{\mathrm{L}}^{S}_{j}}{2}   \right]
\end{equation}
where the Symmetric Logarithmic Derivative (SLD) operators $\hat{\mathrm{L}}^{S}_j$ are implicitly given by the solution of the following Lyapunov Equation \cite{PARIS2009}:
\begin{equation}
    \label{eq:SLDmultiparam}
    \partial_{\lambda_{j}} \rrho_{\vec{\lambda}} \ = \ \dfrac{ \hat{\mathrm{L}}^{S}_{j} \rrho_{\vec{\lambda}} + \rrho_{\vec{\lambda}} \hat{\mathrm{L}}^{S}_{j}}{2}
\end{equation}
The Quantum Fisher Information always results in a valid Cramér-Rao bound (SLD-QCR bound), i.e., 
\begin{equation}
    \label{eq:qCramerRao}
    \Sigma \left( \Theta^{(M)} \right) \ \ \geq \frac{1}{M} \ \mathcal{Q}^{-1} [  \vec{\lambda}^{*}]\,.
\end{equation}
At a variance with the single-parameter case, though, the SLD-QCR bound is not tight for multi-parameter models, in general, since there could be no single POVM saturating it. This is also true for, e.g., the Quantum Cramér-Rao bound derived from the right logarithmic derivative. 

To quantify how tight the SLD-QCR Cramér-Rao bound is, the matrix inequality can be recast into a scalar bound by introducing a {\em weight matrix}, i.e., a $n\times n$ semipositive matrix $\bf W$, and taking the trace of Eqs. (\ref{eq:CramerRao}) and (\ref{eq:qCramerRao}). We have \cite{Albarelli2020}
$$\Tr[{\bf W \, \Sigma}] \ge  C_{\mathcal{F}}({\bf W}) 
\qquad \Tr[{\bf W \, \Sigma}]\ge C_{\mathcal{Q}}({\bf W}),$$ 
where:
\begin{equation}
\begin{aligned}
C_{\mathcal{F}} ({\bf W}) \ &= \ M^{-1} \, \Tr[{\bf W} \, \mathcal{F}^{-1}]  \\
C_{\mathcal{Q}} ({\bf W}) \ &= \ M^{-1} \,\Tr[{\bf W}\, \mathcal{Q}^{-1}]
\end{aligned}
\end{equation}

In the single-parameter scenario, the bound (\ref{eq:qCramerRao}) can be achieved by a projective measurement over the SLD eigenstates, while in the multi-parameter setting,
it is not attainable in general, as the SLDs associated with different parameters may not commute with one another. In this case, the {\em most informative bound}, $C_{\rm MI}({\bf W})= M^{-1} \min_{\mathbf{\Pi}} \{\Tr[{\bf W} \, \mathcal{F}^{-1}]\}$, minimized over all possible measurements $\mathbf{\Pi}$, does not coincide with $C_{\mathcal{Q}}({\bf W})$ in the multiparameter case. Another important bound is the Holevo Cramér-Rao (HCR) bound  $C_{\rm H}({\bf W})$\cite{Holevo1977}, the most informative bound
achievable by collective measurements performed on asymptotically many copies of the state encoding the parameters \cite{Albarelli2020, Razavian20}. In turn, we have $\Tr[{\bf W \, \Sigma}] \ge  C_{\mathcal{F}}({\bf W})  \ge  C_{\rm MI}({\bf W}) \ge  C_{\rm H}({\bf W}) \ge C_{\mathcal{Q}}({\bf W})$.  

The Holevo bound $C_{\rm H}({\bf W})$ is usually difficult to evaluate compared to $C_{\mathcal{Q}}({\bf W})$ and therefore the following relation represents a useful tool in characterizing the a multiparameter estimation model
\begin{eqnarray}\label{eq:HolevoHier}
C_{\mathcal{Q}}({\bf W}) \le C_{\rm H}({\bf W}) \le (1+{\cal R}) C_{\mathcal{Q}}({\bf W}) \, ,
\end{eqnarray}
where the {\em quantumness} parameter ${\cal R}$ 
is given by \cite{carollo20,Carollo2019,candeloro21} 
\begin{eqnarray}\label{eq:R}
{\cal R} = \| i \, \mathcal{Q}^{-1} \mathcal{U} \|_\infty\, .
\end{eqnarray}
In the above equation, $\|{\bf A}\|_\infty$ denotes the largest eigenvalue 
of the matrix $\bf A$, and
$\mathcal{U} (\bf \lambda)$ is the asymptotic incompatibility matrix, also referred to as {\em Uhlmann curvature}, with matrix elements \cite{carollo20}:
\begin{eqnarray}\label{eq:Uhl}
{\mathcal{U}}_{\mu \nu}= -\frac{i}{2} \Tr \left\{\rho_{\bf\lambda} [\hat{\mathrm{L}}_\mu, \hat{\mathrm{L}}_\nu] \right\} \, ,
\end{eqnarray}
where $[A,B]=AB-BA$ is the commutator of $A$ and $B$
{Equation~(\ref{eq:HolevoHier}) implies that QFI bound may be saturated iff ${\mathcal{U}} (\bf\lambda)=0$, which is usually referred to as the {\em weak compatibility condition}.}
The quantumness parameter ${\cal R}$ is bounded $0\le {\cal R} \le 1$ and vanishes ${\cal R}=0$ iff $\mathcal{U} (\bf\lambda)=0$. Therefore, it provides a measure of asymptotic incompatibility between the parameters. For $n=2$ we may write
\begin{eqnarray}\label{eq:R1}
{\cal R} = \sqrt{\frac{\det \mathcal{U}}{\det \mathcal{Q}}} \qquad \mbox{for $n=2$ parameters.}
\end{eqnarray}

\section{Pure-state and unitary quantum statistical models}
For a pure state quantum statistical model with $n$ parameters, $\{ \vert \tilde{\psi} ( \vec{\lambda} )\rangle \}_{\vec{\lambda} \in \reals^n}$, the SLDI-QFI reduces to:
\begin{equation}
    \left[ \QFI \right]_{\mu \nu} \ = \ 4 \mathrm{Re} \left[ \langle \partial_\nu  \tilde{\psi}  \vert \partial_\mu \tilde{\psi} \rangle  -   \langle   \tilde{\psi}  \vert \partial_\mu \tilde{\psi} \rangle   \langle \partial_\nu  \tilde{\psi}  \vert \tilde{\psi} \rangle   \right]
\end{equation}
where $\mu, \nu \in \{ 1,..., n \}$ are indices for the parameters in the vector $\vec{\lambda}$ and having introduced the short-hand notation $ \vert \tilde{\psi} \rangle  \equiv \vert \tilde{\psi} ( \vec{\lambda} )\rangle$. 
If the model is unitary and covariant, meaning that:
\begin{equation}
\label{eq:suppl:unitary}
    \vert \tilde{\psi} ( \vec{\lambda} )\rangle  \ = \ \exp \left[ - i \vec{\lambda} \cdot \vec{\hat{G}}   \right] \vert \psi_0 \rangle
\end{equation}
where $\vec{\hat{G}}$ is a vector of $n$ self-adjoint operators that are independent of $\vec{\lambda}$, then one can easily show that \cite{PARIS2009}:
\begin{equation}
     \left[ \QFI \right]_{\mu \nu}  \ \ = \ \ 4 \  \mathrm{Cov}_{\psi_0} \left(  \hat{G}_\mu, \hat{G}_\nu \right)
\end{equation}
where:
\begin{equation}
     \mathrm{Cov}_{\psi_0} \left(  \hat{G}_\mu, \hat{G}_\nu \right)  \ \  := \ \ \Big\langle \dfrac{\hat{G}_\mu \hat{G}_\nu + \hat{G}_\nu \hat{G}_\mu  }{2} \Big\rangle_{\psi_0} \ - \ \big\langle \hat{G}_\mu \big\rangle_{\psi_0} \big\langle \hat{G}_\nu \big\rangle_{\psi_0}
\end{equation}
and $\langle \hat{A} \rangle_{\psi_0} = \langle \psi_0 \vert \hat{A} \vert \psi_0 \rangle$. In particular, the SLD-QFI matrix is independent of $\vec{\lambda}$. \\

Still considering the pure state model and introducing the notation $\hat{\varrho}_{\lambda} = \vert \tilde{\psi}(\vec{\lambda})\rangle\langle\tilde{\psi}(\vec{\lambda})\vert$, we can compute the symmetric logarithmic derivative by using the fact that:
\begin{equation}
 \frac{1}{2} \left[ \hat{\mathrm{L}}_\mu , \hat{\varrho}_\lambda \right] \ = \ \partial_\mu \hat{\varrho}_{\lambda} \ = \ \partial_\mu \hat{\varrho}^{2}_{\lambda} \ = \ 2 ( \partial_\mu \hat{\varrho}_{\lambda} ) \hat{\varrho}_{\lambda} + 2 \hat{\varrho}_{\lambda} ( \partial_\mu \hat{\varrho}_{\lambda}) 
\end{equation}
from which it follows that:
\begin{equation}
    \hat{\mathrm{L}}_{\mu} \ = \ 2 \partial_\mu \hat{\varrho}_{\lambda} \ = \ 2 \big( \vert \partial_{\mu} \tilde{\psi}\rangle\langle\tilde{\psi}\vert + \vert \tilde{\psi}\rangle\langle \partial_{\mu} \tilde{\psi}\vert \big)
\end{equation}
and in the case of the unitary model of Eq.(\ref{eq:suppl:unitary}):
\begin{equation}
    \hat{\mathrm{L}}_{\mu} \ = \  -2 i \big( \hat{G}_{\mu} \vert  \tilde{\psi} \rangle\langle \tilde{\psi} \vert - \vert \tilde{\psi} \rangle\langle \tilde{\psi} \vert  \hat{G}_{\mu} \big)
\end{equation}
Since the variance of $\hat{G}_{\mu}$ on $\vert \tilde{\psi} \rangle$ is proportional to the SLD-QFI for the parameter $\lambda_{\mu}$, we should not pick $\vert \tilde{\psi} \rangle$ to be an eigenvalue of $\hat{G}_{\mu}$. This means that $ \{ \vert \tilde{\psi} \rangle, \hat{G}_{\mu} \vert \tilde{\psi}  \rangle \}$ are linearly independent. The normalized component of $\hat{G}_\mu \vert \tilde{\psi} \rangle$ orthogonal to $ \vert \tilde{\psi} \rangle$ is given by:
\begin{equation}
\label{eq:suppl:SLD1}
\vert \psi_{1} \rangle  \ \ := \ \ \dfrac{ \hat{G}_{\mu} - \langle \hat{G}_{\mu} \rangle_{\psi_0}}{\sqrt{\Delta^{2}_{\psi_0} \hat{G}_{\mu}}} \vert \tilde{\psi} \rangle
\end{equation}
where $(\Delta^{2}_{\psi_0} \hat{G}_{\mu}) = \mathrm{Cov}_{\psi_0}(\hat{G}_{\mu}, \hat{G}_{\mu})$. In the relevant 2-dimensional subspace spanned by the orthonormal basis $\{ \vert \tilde{\psi} \rangle, \vert \psi_1 \rangle \}$, then, the symmetric logarithmic derivative is represented by the matrix:
\begin{equation}
\label{eq:suppl:SLD2}
    \hat{\mathrm{L}}_{\mu} \ = \  2 \sqrt{\Delta^{2}_{\psi_0} \hat{G}_{\mu}} \left(  \begin{array}{cc}  0 & - i \\ i & 0 \end{array}  \right)
\end{equation} 
whose eigenvalues are $\pm 2 \sqrt{\Delta^{2}_{\psi_0} \hat{G}_{\mu}}$, while the corresponding eigenvectors, which \emph{define the projective measurement to be performed to compute the SLD-QFI for the parameter $\lambda_{\mu}$}, are $\frac{\vert \tilde{\psi} \rangle \pm i \vert \psi_{1} \rangle }{\sqrt{2}}$. This means that one has to project on a superposition of $\vert \tilde{\psi} \rangle$ and the state $\hat{G}_{\mu} \vert \tilde{\psi} \rangle$. It is important to stress, however, that in the multi-parameter scenario the SLD operators do not automatically define the best POVM to be measured to simultaneously estimate all the parameters with the highest possible precision: indeed, in our case and as it often happens, they do not even commute and they do not fit in a single POVM. Rather, having shown that for suitably chosen probe states we can have a vanishing R-parameter, this means that the SLD Cramér-Rao bound coincides with the Holevo bound, which is in general attainable only with some \emph{collective measurement protocol} on asymptotically many copies of the single-mode state encoding the parameters.

\section{Estimation of two incompatible displacement parameters with Fock states}
\label{appx:Fock}

As outlined in the main text, we will now consider the scenario with $2$ parameters and $2$ generators, $\hat{G}_1 = \hat{p}$ corresponding to $\lambda_1 = x_0$ and $\hat{G}_2 = \hat{x}$ with $\lambda_2 = p_0$. Explicitly, the pure quantum statistical model is defined as: 
\begin{equation}
    \vert \tilde{\psi} (x_0, p_0) \rangle \ = \ e^{- i p_{0} \hat{x}} e^{- i x_{0} \hat{p}} \vert \psi \rangle
\end{equation}
Notice that this is \emph{not} in the form of Eq.(\ref{eq:suppl:unitary}). However, we have that $\ e^{- i p_{0} \hat{x}} e^{- i x_{0} \hat{p}} = \ e^{i \frac{x_0 p_0}{2}} e^{- i p_{0} \hat{x}- i x_{0} \hat{p}}$, therefore it can be cast in that form apart for an overall constant phase which is immaterial. 
Recall that the so-called photon number operator $\hat{n} = \hat{a}^{\dagger}\hat{a}$ is equal to $\hat{n} = \frac{\hat{x}^2 + \hat{p}^2 - 1}{2}$. Taking the expectation value on the probe state $\vert \psi_0 \rangle$ and calling $E = \langle \psi_0 \vert \hat{n} \vert \psi_0 \rangle$ its average energy ($\hbar = \omega = 1$), we find that:
\[  E \ = \ \frac{ \langle \hat{x}^2 \rangle_{\psi_0}  +  \langle \hat{p}^2 \rangle_{\psi_0} - 1}{2}  \ \geq \ \frac{\Delta^{2}_{\psi_0} \hat{x}+\Delta^{2}_{\psi_0} \hat{p}}{2} - \frac{1}{2}      \]
from which follows that:
\begin{equation}
    \Delta^{2}_{\psi_0} \hat{x} \ \leq \ 2E + 1 \ \ , \ \ \ \   \Delta^{2}_{\psi_0} \hat{p} \ \leq \ 2E + 1
\end{equation}
proving that both variances cannot grow more than linearly in the energy of the probe state, thus preventing an Heisenberg limit type of scaling. 
Without loss of generality, we can also assume $ \Delta^{2}_{\psi_0} \hat{x} =  \Delta^{2}_{\psi_0} \hat{p}$, since we are interested in estimating both parameters simultaneously and there is no reason to introduce unbalances, which further reduces the inequality to:
\begin{equation}
    \Delta^{2}_{\psi_0} \hat{x} \ =  \Delta^{2}_{\psi_0} \hat{p} \ \leq \ E + \frac{1}{2}
\end{equation}
which is saturated by Fock states. Therefore, Fock states are optimal probe states at fixed energy to simultaneously estimate $x_0$ and $p_0$ with an arbitrarily high precision and compatibility. \\

\section{Every covariance matrix can be ``purified'' by a non-Gaussian state}

Here we show that for any matrix $\sigma$ that satisfies the Heisenberg uncertainty relation one can find a non-Gaussian pure state $\lvert\psi_{\sigma}\rangle$ with $\sigma$ as its covariance matrix. In this sense, one could say that $\lvert\psi_{\sigma}\rangle$ is a ``purification'' of the covariance matrix $\sigma$.

As a proof, notice first of all that any single-mode $\sigma$ that satisfies the uncertainty relation can be written, using the Williamson decomposition, as
\begin{equation}
    \sigma = S^{\top}\sigma_T S,
\end{equation}
where $S$ is a symplectic matrix that implements a Gaussian unitary transformation $\hat U_S$ on the state, and $\sigma_T$ is the covariance matrix of a thermal state. It now suffices to show that for every $\sigma_T$, we can find a pure state $\lvert\psi_{\sigma_T}\rangle$ with $\sigma_T$ as quadrature covariance matrix. Subsequently we can identify $\lvert\psi_{\sigma}\rangle = \hat U_S \lvert\psi_{\sigma_T}\rangle$

First, we assume that we are dealing with a single mode, such that $\sigma_T = n \mathds{1}_2$, with $n \geq 1/2$. We can now define 
\begin{equation}
\lvert\tau\rangle = \sqrt{\lambda} \lvert N\rangle + \sqrt{1-\lambda} \lvert 0\rangle,
\end{equation}
where we choose the Fock state $\lvert N\rangle$ such that $N \geq \max(2,n)$. We find that the covariance matrix of this state is given by
\begin{equation}
    \sigma_{\tau} = (\lambda N + \frac{1}{2}) \mathds{1}.
\end{equation}
It now suffices to set $\lambda = (2n - 1)/2N$ to find that
\begin{equation}
    \sigma_{\tau} = \sigma_T.
\end{equation}
It should be noted that this choice of $\lvert\tau\rangle$ is highly non-unique. 

Finally, we highlight that this result can very easily be extended to a multimode setting, given that for an $m$-mode system the covariance matrix $\sigma_T$ is given by
\begin{equation}
    \sigma_T = \bigoplus_{k=1}^m  n_k \mathds{1}_2.
\end{equation}
We can then simply find
\begin{equation}
\lvert\psi_{\sigma_T}\rangle = \lvert\psi_{\tau_1}\rangle\otimes\dots \otimes\lvert\psi_{\tau_m}\rangle,
\end{equation}
where every $\lvert\psi_{\tau_k}\rangle$ is given by
\begin{equation}
\lvert\psi_{\tau_k}\rangle = \sqrt{\lambda_k} \lvert N_k\rangle + \sqrt{1-\lambda_k} \lvert 0\rangle,
\end{equation}
where we set $\lambda_k = (2n_k - 1)/2N_k$, and again fix $N_k \geq \max(2,n_k)$. The then find the overall state
\begin{equation}
    \lvert\psi_{\sigma}\rangle = \hat U_S (\lvert\psi_{\tau_1}\rangle\otimes\dots \otimes\lvert\psi_{\tau_m}\rangle),
\end{equation}
where it should be noted that $\hat U_S$ is in general a multimode Gaussian unitary (or Bogoliubov transformation for those who prefer that terminology) that can (and generally will) create entanglement.